\documentclass[letterpaper,twocolumn,10pt]{article}
\usepackage{usenix2019_v3}
\usepackage[utf8]{inputenc} 
\usepackage[T1]{fontenc}    
\usepackage{hyperref}       
\usepackage{url}            
\usepackage{booktabs}       
\usepackage{amsfonts}       
\usepackage{nicefrac}       
\usepackage{microtype}      
\usepackage{xcolor}         
\usepackage{amsmath}
\usepackage{dsfont}
\usepackage{amsthm}
\usepackage{multirow}
\usepackage{makecell}
\usepackage{subfigure}
\usepackage{wrapfig}
\usepackage{graphicx}
\usepackage{algorithm}
\usepackage{algorithmic}
\usepackage{CJKutf8}
\usepackage{tikz}
\usepackage{amsmath}

\newtheorem{definition}{Definition}

\usepackage{filecontents}
\usepackage{tcolorbox} 
\usepackage{varwidth} 
\tcbuselibrary{breakable} 
\tcbuselibrary{skins} 
\usepackage{float} 
\usepackage[available]{usenixbadges}
\usepackage{graphicx}
\usepackage{tabularx}
\usepackage{subfigure}
\usepackage{amsmath, amssymb}
\usepackage{booktabs}
\usepackage{multirow}
\usepackage{makecell}
\usepackage{algorithm}
\usepackage{algorithmic}
\usepackage[table]{xcolor} 
\usepackage{booktabs}
\usepackage{tabularx}
\usepackage{array}      
\usepackage{longtable}
\newcolumntype{L}{>{\raggedright\arraybackslash}X} 
\makeatletter
\renewcommand\thanks[1]{%
  \protected@xdef\@thanks{\@thanks
    \protect\footnotetext[\the\c@footnote]{#1}}%
  \stepcounter{footnote}}
\makeatother

\NewTColorBox{theobox}{o m +o}{
    enhanced,
    colframe=gray!0!white,
    interior empty,
    coltitle=white,
    fonttitle=\mdseries,
    colbacktitle=gray,
    extras broken={frame empty,interior empty},
    borderline={0.5mm}{0mm}{gray},
    sharp corners=downhill,
    top=4mm,
    before skip=3.5mm,
    attach boxed title to top left={yshift=-3mm,xshift=5mm},
    boxed title style={boxrule=0pt,sharp corners=all},varwidth boxed title,
    IfNoValueTF={#1}{title=#2}{title=#2:~#1},
    IfNoValueTF={#3}{}{#3}
}

\begin{document}

\date{}

\title{\Large \bf Provable Secure Steganography Based on Adaptive
Dynamic Sampling }

\author{
{\rm Kaiyi Pang,  Minhao Bai}
\\
Tsinghua University
}


\maketitle

\begin{abstract}
The security of private communication is increasingly at risk due to widespread surveillance. Steganography, a technique for embedding secret messages within innocuous carriers, enables covert communication over monitored channels. Provably Secure Steganography (PSS), which ensures computational indistinguishability between the normal model output and steganography output, is the state-of-the-art in this field. However, current PSS methods often require obtaining the explicit distributions of the model.
In this paper, we propose a provably secure steganography scheme that only requires a model API that accepts a seed as input. Our core mechanism involves sampling a candidate set of tokens and constructing a map from possible message bit strings to these tokens. The output token is selected by applying this mapping to the real secret message, which provably preserves the original model's distribution. To ensure correct decoding, we address collision cases, where multiple candidate messages map to the same token, by maintaining and strategically expanding a dynamic collision set within a bounded size range.
Extensive evaluations of three real-world datasets and three large language models demonstrate that our sampling-based method is comparable with existing PSS methods in efficiency and capacity.
\end{abstract}

\section{Introduction}

Steganography is a technique for embedding and transmitting private messages within seemingly innocuous carriers, such as images or text, over monitored channels \cite{imagestega1, imagestega2, meteor2021, ding2023discop}. Its primary goal is to conceal not only the content of secret messages but also the fact that secret communication is taking place. A normal carrier is referred to as a cover, while a carrier that contains hidden information but still appears normal is called a stego.

Steganography methods can be roughly categorized into modification-based \cite{xiugai1,xiugai2}, retrieval-based \cite{jiansuo1,jiansuo2}, and generation-based.
With the rapid advancement of generative models capable of approximating target data distributions and synthesizing realistic carriers (e.g., ChatGPT generating human-like text or Diffusion models creating high-quality images), generative steganography has become a popular approach~\cite{yang2018rnn,ding2023discop,meteor2021}. It eliminates the need for pre-existing cover carriers and significantly boosts capacity by leveraging the explicit distributions produced by the generative models.
Many researchers have developed advanced steganography techniques for mapping the distributions produced by language models to secret messages, such as Bins \cite{fang2017generating}, Arithmetic Coding (AC) \cite{ziegler2019neural}, Huffman Coding (HC) \cite{yang2018rnn}, and Adaptive Dynamic Grouping (ADG) \cite{zhang2021provably}, enabling the generation of fluent stegotext with high capacity. However, these heuristic methods often alter the original generative model’s distribution, introducing security risks. Well-designed steganalysis classifiers \cite{xue2023adaptive,TS-CSW} can exploit this feature, gaining a noticeable advantage in distinguishing between cover and stego distributions.

Researchers have begun exploring provably secure steganography since 2 decades ago \cite{cachin1998information,hopper2004toward}. In 1998, Cachin \cite{cachin1998information} modeled steganography systems from an information-theoretic perspective, using Kullback–Leibler (KL) divergence between cover and stego distributions to evaluate security. From a computational complexity perspective, Hooper \cite{hopper2004toward} laid the groundwork for provably secure steganography. A stegosystem is considered secure if any probabilistic polynomial-time distinguisher cannot effectively differentiate between the distributions of stego and cover.

Recent advancements in generative models have enabled the simulation and generation of realistic carrier distributions, driving the practical application of provably secure steganography. For textual carriers, current Large Language Models (LLMs) can produce a precise distribution over the next token. State-of-the-art methods like METEOR \cite{meteor2021}, DISCOP \cite{ding2023discop}, iMEC \cite{de2022perfectly}, and SparSamp\cite{wang2025sparsamp}, generate secure stegos by manipulating the sampling process based on the model distributions and the secret messages. 

However, current provably secure methods depend entirely on the explicit distributions provided by generative models. In real-world scenarios, Many high-quality generative models do not expose their output distributions, for example, because providers want to protect intellectual property and prevent model distillation. Instead, they only provide the output samples (such as texts). This lack of access to model distributions makes it impossible to apply these provably secure methods \cite{meteor2021,ding2023discop,wang2025sparsamp} in scenarios where the model’s distribution is unknown.

In this paper, we propose a provably secure steganography method that does not rely on explicit model distributions, in contrast to existing PSS methods that require them~\cite{ding2023discop,IMECde2022perfectly,wang2025sparsamp,meteor2021,liao2025framework}. Our approach operates effectively using a model API that accepts a seed as input and outputs only tokens. Building upon the METEOR scheme~\cite{meteor2021}, we adapt its principles to the setting of API-only access. As METEOR encodes secret bits by converting them into a decimal value to perform inverse transform sampling, it becomes infeasible in API-only scenarios due to the absence of distribution. Our method overcomes this limitation, enabling secure steganography under practical constraints.

Our approach is based on two insights. First, by sampling multiple candidate tokens from the model API and selecting one uniformly, the output distribution should be the same as the model's prediction. This allows constructing a mapping $F$ from all possible $N$-bit strings to sampled tokens, and choosing the token by applying the mapping $F$ to the secret message. Second, collisions, where different message strings are mapped to the same token, do not hinder embedding but complicate decoding. Instead of avoiding collisions as METEOR, we maintain a list of all colliding strings (including the true secret message). In subsequent steps, it is feasible to sample as many candidates as there are elements in the collision set, which typically shrinks to size 1 after a few steps. To prevent capacity loss from prolonged small collision set sizes, we dynamically expand the list by appending `0` and `1` to each element, doubling its size until a predefined limit (e.g., $2^n$) is reached. This bounds the size of the collision set between $2^{n-1}$ and $2^n$, ensuring sufficient samples for high capacity. At the end of the text, we extract the shared prefix of the colliding strings, minimizing the negative impact of frequently doing prefix extraction which reduces capacity.

Basically, our steganography scheme has 3 main differences from the METEOR scheme. First is that we draw samples from a model API rather than get access to the model distribution. Directly applying METEOR to model APIs will require sampling for $2^{32}$ or more times, which is not feasible. While in our scheme, we can sample for $2^4$ or fewer times but still maintain the security. Second, our scheme maintains a list of possible candidate messages during the encoding process, which avoids the loss of capacity introduced by frequently extracting the shared prefix. Third, the number of samples in our scheme can vary, and dynamically varying the number of samples in response to the current number of candidate messages can allow higher throughput.

The main contributions of this paper are as follows:  
\begin{itemize}
    \item We propose a provably secure steganography scheme based on making queries to an appropriate model API with seed support and provide a formal security proof.
    \item We propose a sampling strategy that maintains a list of possible candidate secret messages at any intermediate step, as long as the number of candidates does not exceed a certain limit. The number of samples in each step dynamically varies in response to the current number of candidate messages and can allow higher throughput.
    \item Extensive experiments show that our scheme is both effective and efficient. Without access to the explicit model’s distribution, our method achieves comparable or even superior embedding capacity to methods that require explicit distribution. 
\end{itemize}

\section{Related Work}
\begin{table}[]
	\centering
 {
 {\color{black}
	\begin{tabular}{cccc}\toprule[1.5pt]
		Method   &Mode & Complexity & Correct?  \\\midrule
		METEOR~\cite{meteor2021}  &     Stream                     &      $O(|V|)$    &  100\% \\
  METEOR(R.)~\cite{meteor2021}  &     Stream                     &      $O(2^{|H(\mathcal{P})|}|V|)$    &  100\% \\
		DISCOP~\cite{ding2023discop}   &       Stream                   & $O(|V|)$         &  100\%  \\
  DISCOP(R.)~\cite{ding2023discop}   &       Stream                   & $O(|V|log|V|)$         &  100\%  \\
		iMEC~\cite{de2022perfectly}     &           Block                &    $O(2^{B}|V|)$     & Unstable  \\
		SparSamp~\cite{wang2025sparsamp} &       Block                    &   $O(2^{B})$ &  100\% \\ 
		Ours     &          Stream                &      $O(2^{N})$ & 100\% \\\bottomrule[1.5pt]
	\end{tabular}}}
	\caption{Comparison of provable secure steganography schemes, where the $(R.)$ represents the reordered version. $|V|$ represents the size of model vocabulary, typically between $2^{17}\sim 2^{19}$ for LLM. $H(\mathcal{P})$ represents the entropy of the model distribution, typically between $0.91 \sim 1.05$ for model \textsc{Mistral}. $B$ represents the size of a block, which can vary between $8 \sim 64$. $2^N$ is the maximum number of samples in our scheme.  $2^N$ can vary from $2^2 \sim 2^{15}$. }
	\label{tab:summary}
\end{table}
\subsection{Definition of Security}
There are two definitions of security in steganography. 
Cachin \cite{cachin1998information} first modeled steganographic security by the Kullback-Leibler Divergence (KL-Divergence) between the stegotext and covertext:
\begin{equation}
    D_{KL}(P_S||P_C) = \sum_x P_S(x)\log\frac{P_S(x)}{P_C(x)} < \epsilon,
\end{equation}
where $D_{KL}(P_S||P_C) = 0$ denotes perfect security.

Different from the Cachin, another widely used steganography security is based on the complexity theory in the perspective of \textit{chosen hiddentext attacks} \cite{hopper2004toward}. It requires that for all probabilistic polynomial-time (PPT) adversaries $\mathcal{A}$, $k \leftarrow \mathsf{KeyGen}(1^\lambda)$, the probability that $\mathcal{A}$ can effectively distinguish between stego and cover is negligible:
\begin{align}
    &|\text{Pr}[\mathcal{A}^{{\mathsf{Encode}}_k(\cdot,\cdot)} = 1] - \text{Pr}[\mathcal{A}^{\mathsf{Model}(\cdot,\cdot)} = 1]| < \mathsf{negl}(\lambda),
\end{align}
where $\mathsf{Model}$ denotes the generative model, ${\mathsf{Encode}}$ represents the steganography encoding algorithm, and $\mathsf{negl}(\lambda)$ is a negligible function.

\subsection{Entropy Coding Based Steganography}

Baron et al. \cite{Baron03} were the first to introduce the duality between information embedding and source coding. Building on this connection, Ziegler et al. \cite{ziegler2019neural} explored the use of arithmetic coding (AC) for steganography, while Satir et al. \cite{Satir14} and Yang et al. \cite{yang2018rnn} applied Huffman coding (HC). These methods leverage entropy coding to assign codewords to symbols based on their probabilities. As the codewords are prefix-free, secret bits can be matched to a single codeword, and the corresponding symbol transmitted. 
However, these entropy-coding-based steganographic schemes raise security concerns. Ziegler’s construction~\cite{ziegler2019neural} suffers from a randomness-reuse issue, as highlighted by Kaptchuk~\cite{meteor2021}. HC-based approaches suffer from mismatched probabilities between codewords and symbols, as symbol probabilities are not always negative integral powers of 2. This causes the generated stego to be more detectable.
Despite these limitations, entropy coding represents the theoretical upper limit for channel entropy utilization. With improved security measures, the focus will shift towards maximizing the steganographic embedding rate, with entropy coding offering valuable insights for future research.

\subsection{METEOR}
Kaptchuk et al.\cite{meteor2021} proposed METEOR, a secure steganography that re-encrypts the embedded bits step by step. This work points out that previous methods \cite{yang2018rnn,ziegler2019neural} have the problem of randomness reuse, which will cause the leakage of hidden information. METEOR collects several tokens until the entropy is enough for embedding, and re-encrypts the rest of the bits to avoid information leakage. This method is computationally secure, but the entropy is not fully used. After each embedding, the secret bits are re-encrypted, which effectively truncates the coding process and wastes entropy at the end of every embedding step. When the model distribution has very low entropy (as is often the case for LLMs), there may be no common prefix at all, leading to low embedding capacity. In order to further improve the expectation of embedding capacity, an reorder algorithm is proposed for METEOR. The computational complexities of the original and reordered algorithms are $O(|V|)$ and $O(2^{|H(\mathcal{P})|}\,|V|)$, respectively, where $|V|$ is the model’s vocabulary size and $H(\mathcal{P})$ is the entropy of the model distribution.

\subsection{Distribution Copies}
Ding et al.\cite{ding2023discop} proposed a provably secure steganography method based on copies of the model distribution. In the situation that the random numbers $r \in [0,1)$ and $r + \frac{1}{2} \mod 1$ point to different tokens in the copies, these 2 special tokens can be chosen according to the secret bit $0$ or $1$. If the random numbers $r$, $r + \frac{1}{2^n} \mod 1$, $r + \frac{2}{2^n} \mod 1$, ... , $r + \frac{2^n -1}{2^n} \mod 1$ point to different tokens, the embedding capacity is $n$ bits. Since repeated copies of the distribution can exhibit substantial overlap, the embedding rate is severely constrained and becomes asymptotically limited by the minimum entropy. To further increase capacity, they proposed a Huffman-tree–based variant. Considering the typical case of constructing a Huffman tree, the computational complexity is $O(|V|\log |V|)$, where $|V|$ denotes the vocabulary size of the model.

\subsection{Minimum Entropy Coupling (MEC)}
Minimum Entropy Coupling (MEC) is the problem that given the marginal distribution of the random variables, computing the joint distribution that has the minimum entropy. The core of this problem is to find how relevant it is likely to be between these variables. More details can be found in the excellent work of Cicalese et al. \cite{Cicalese19}. De Witt et al.\cite{de2022perfectly} first explore using MEC to construct secure and efficient steganography. They try to compute MEC matrix between the uniform distributed $n$-bit codewords and the distribution predicted by language models. Then they choose the $n$-bit prefix of secret messages as the chosen codeword, and randomly sample a token from the marginal distribution of this codeword. Usually, the $n$-bit codewords cannot be embedded in one loop; this process will be repeated several times until the bits can be uniquely decoded. However, computing the minimum entropy coupling is an NP-hard problem. Many efforts have been made to design a poly-time algorithm to asymptotically approach the MEC \cite{kocaoglu2017entropic,Cicalese19,Kocaoglu17,Li21}.
MEC-based steganography has 2 weaknesses: (i) up to now, the best MEC algorithm is only able to compute a joint distribution whose entropy is within 1 bit of MEC, resulting in sustained loss of capacity and (ii) the complexity is at least $O(n\text{log}n)$\cite{Cicalese19}. Therefore, the large set of codewords will make this type of steganography inefficient. Moreover, practical deployment exposes additional issues, such as decoding errors and potential dead cycles. Since iMEC relies on probabilistic decoding, it cannot provide a deterministic guarantee on the decoding error rate.
MEC-based steganography still needs more exploration.

\subsection{SparSamp}
Wang et al proposed SparSamp~\cite{wang2025sparsamp}, a provably secure steganographic algorithm that represents the latest advancement in provable security within the field. Based on DISCOP~\cite{ding2023discop}, the embedding capacity is improved by increasing the spacing between sampling intervals and cumulating entropy, without introducing additional computational complexity. Although SparSamp is claimed to achieve $O(1)$ complexity, we cautiously suggest that in practice, its effective complexity may still depend on the choice of block size $B$. However, this method still requires both the sender and receiver to share the same model distribution. Moreover, due to the fixed-length embedding of secret information, the approach does not fully exploit the entropy of the cover text, leaving room for further improvement in embedding capacity.

\section{Method}

\begin{figure*}[htb]
	\centering
	\includegraphics[width=0.8\linewidth]{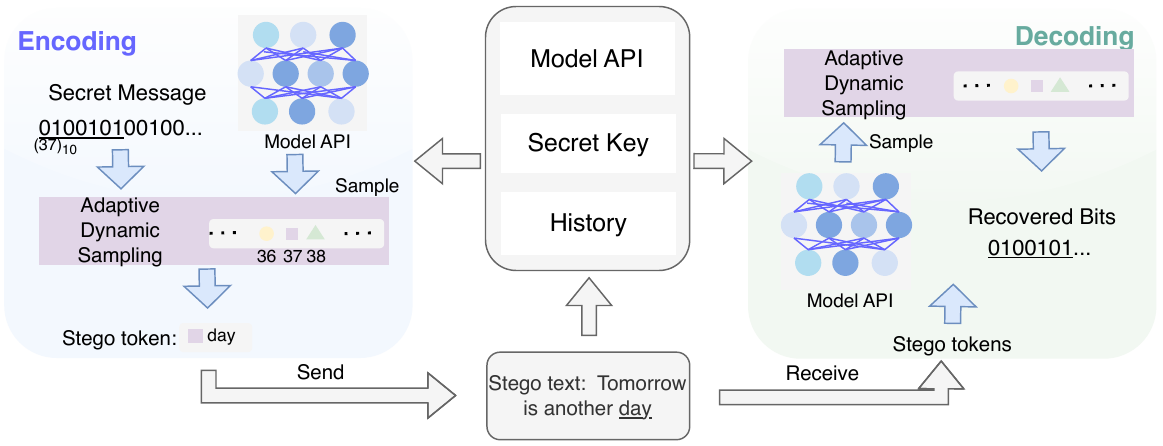}
	\caption{Overall framework of covert communication. }
	\label{overall}
\end{figure*}

Figure~\ref{overall} presents the overall architecture of covert communication. The sender and receiver share a secret key, the history, and a model API that accepts a seed as input. Given the secret message, the sender applies the encoding algorithm by querying the model API multiple times to generate samples, and then selects the sample corresponding to the secret message as output. The stegotext is generated autoregressively until  the sentence naturally terminates. The decoding algorithm executes the inverse procedure, given the stegotext, it performs the same sampling procedure over the model API to recover the secret message.

\subsection{Notations}
In this paper, we mainly use the following notations:
\begin{itemize}
    \item \textbf{Parameters.} We use $k$ to represent the private key and $\lambda$ to denote the security parameter.
    \item \textbf{Sets.} We use $\{0,1\}^N$ to denote the set of all $N$-bit strings and $\{0,1\}$ to denote the bit strings with finite length.
    \item \textbf{Adversaries.} We use $\mathcal{A}$ to denote the adversaries. In most of the paper, the adversary is often written as $\mathcal{A}^{\mathsf{Oracle}}(1^\lambda)$. Here the superscript $\mathsf{Oracle}$ means that the adversary can query the $\mathsf{Oracle}$ for at most polynomial times, and $1^\lambda$ is the internal randomness of the adversary.
    \item \textbf{Algorithms.} We typically write algorithm names in a $\mathsf{\textsf{}}$ font. For instance, the encoding algorithm is denoted as $\mathsf{Enc}_{\mathsf{Model}, k}(m, h)$. Here, the subscripts $\mathsf{Model}$ and the key $k$ are hyperparameters of the algorithm, whereas the message $m$ and the history $h$ are its input.
\end{itemize}
\subsection{Prelimilary}

According to Hopper's notion \cite{hopper2004toward}, steganography schemes should be based on some efficiently sampleable channels. As we use the model API to get samples, we formally define the sampleable channel $\mathcal{C}$ as follows.
\begin{definition}[Sampleable channel]
    A sampleable channel $\mathcal{C}_{k}(h)\xrightarrow{} t$ is a keyed function that outputs a unit of covertext $t$ based on some history $h$.
\end{definition}
So if the sender and the receiver share a secret key, they can get the same samples from the sampleable channel $\mathcal{C}_{k}(h)$. As we use model API to get the samples, in the rest of this paper, we often write the channel as $\mathsf{Model}_{seed}(h)$.  Here we formally describe the steganography scheme.
\begin{definition}[Steganography scheme]
    A steganography scheme $\Sigma_{\mathsf{Model}}$ is a triple of probabilistic algorithms $\Sigma_{\mathsf{Model}}=( \mathsf{KeyGen},  \mathsf{Enc}_{\mathsf{Model},k}, \mathsf{Dec}_{\mathsf{Model},k})$.
    \begin{itemize}
    \item $\mathsf{KeyGen}(1^\lambda)$ is a randomized algorithm that takes an arbitrary input of length $\lambda$ and generates the key $k$ shared between the sender and receiver.
    \item $\mathsf{Enc}_{\mathsf{Model},k}(m,h)$ is a keyed randomized algorithm that takes  a secret message $m \in \{0,1\}^{*}$ and the channel history $h$ as input and outputs a stegotext $s$. 
    \item $\mathsf{Dec}_{\mathsf{Model},k}(s,h)$ is a keyed randomized algorithm that takes as input the key $k$, the stegotext $s$ and the history $h$ and outputs the secret message $m$. 
\end{itemize}
\end{definition}
Basically, a useful steganography scheme $\Sigma_{\mathsf{Model}}$ should satisfy \textit{correctness} and \textit{security}.

\begin{definition}[Correctness]
    We say a steganography scheme $\sum_{\mathsf{Model}}$ is \textit{correct} if for any history $h$ and any message $m$, the stegotext generated by $\mathsf{Enc}_{\mathsf{Model},k}(m,h)$ should be recovered by the decoding algorithm only with negligible probability of error.
    \begin{equation}
\begin{aligned}
    \Pr[\mathsf{Dec}_{\mathsf{Model},k}(\mathsf{Enc}_{\mathsf{Model},k}(m,h),h)= m]\geq 1-\mathsf{negl}(\lambda),
\end{aligned}
\end{equation}
where $\mathsf{negl}(\lambda)$ is the negligible function that correlates to the secure parameter $\lambda$.
\end{definition}

\begin{definition}[Security]
    We say a steganography scheme $\Sigma_{\mathsf{Model}}$ is secure if given the fixed secret message $m$, for any history $h$, for all probabilistic polynomial-time adversary $\mathcal{A}$, he cannot effectively distinguish the stegotext generated by $\mathsf{Enc}$ and the covertext generated by $\mathsf{Model}$.
    \begin{equation}
        \begin{aligned}
            |\Pr[\mathcal{A}^{\mathsf{Enc}_{\mathsf{Model},k}(m,h)}(1^\lambda)=1]-\Pr[\mathcal{A}^{\mathsf{Model}(h)}(1^\lambda)=1]|&\\
            \leq \mathsf{negl}(\lambda).&
        \end{aligned}
    \end{equation}
    Here the $1^\lambda$ represents the internal randomness of adversary $\mathcal{A}$, and $\mathcal{A}$ outputs $1$ if he recognizes the output of $\mathsf{Enc}$ or $\mathsf{Model}$ as stegotext.
\end{definition}

\subsection{Intuition}
We mainly follow the METEOR \cite{meteor2021}, which requires the explicit model distribution for codec. Our goal is to convert the METEOR scheme into a steganography scheme that can work when access is limited to a model API. In METEOR, it uses a binary mask string $b_m$ to XOR with the secret bit string $b_s$, the result $b = b_m \oplus b_s$ is a pseudorandom bit string. Then, the encoding algorithm transforms the string $b$ in to a decimal form $\Bar{b} = \sum_{i = 1} ^ {|b|} b_i \cdot 2^{-i}$, where the $b_i$ is the $i$-th bit of $b$, and $|b|$ is the length of $b$. After that, the encoding algorithm uses the $\Bar{b}$ as the random number to sample from the model distribution. In a different view, we can describe the encoding procedure of METEOR as first generating $2^{|b|}$ samples, then compute the $\Bar{b}$ and choose one sample corresponding to $\Bar{b}$. If we do have the explicit distribution, there is no need to generate $2^{|b|}$ samples as we only have to compute $\Bar{b}$ and sample from the distribution according to $\Bar{b}$. However, if we only have a model API, sampling $2^{|b|}$ times may be too large to be realistic. When using METEOR, the window length of the secret message is often set as $32$ or $64$, which is not suitable for sampling from model API. Therefore, our goal is to decrease the number of samples while ensure the security.

\begin{figure*}[t]
    \centering
     \includegraphics[
        width=0.8\textwidth
    ]{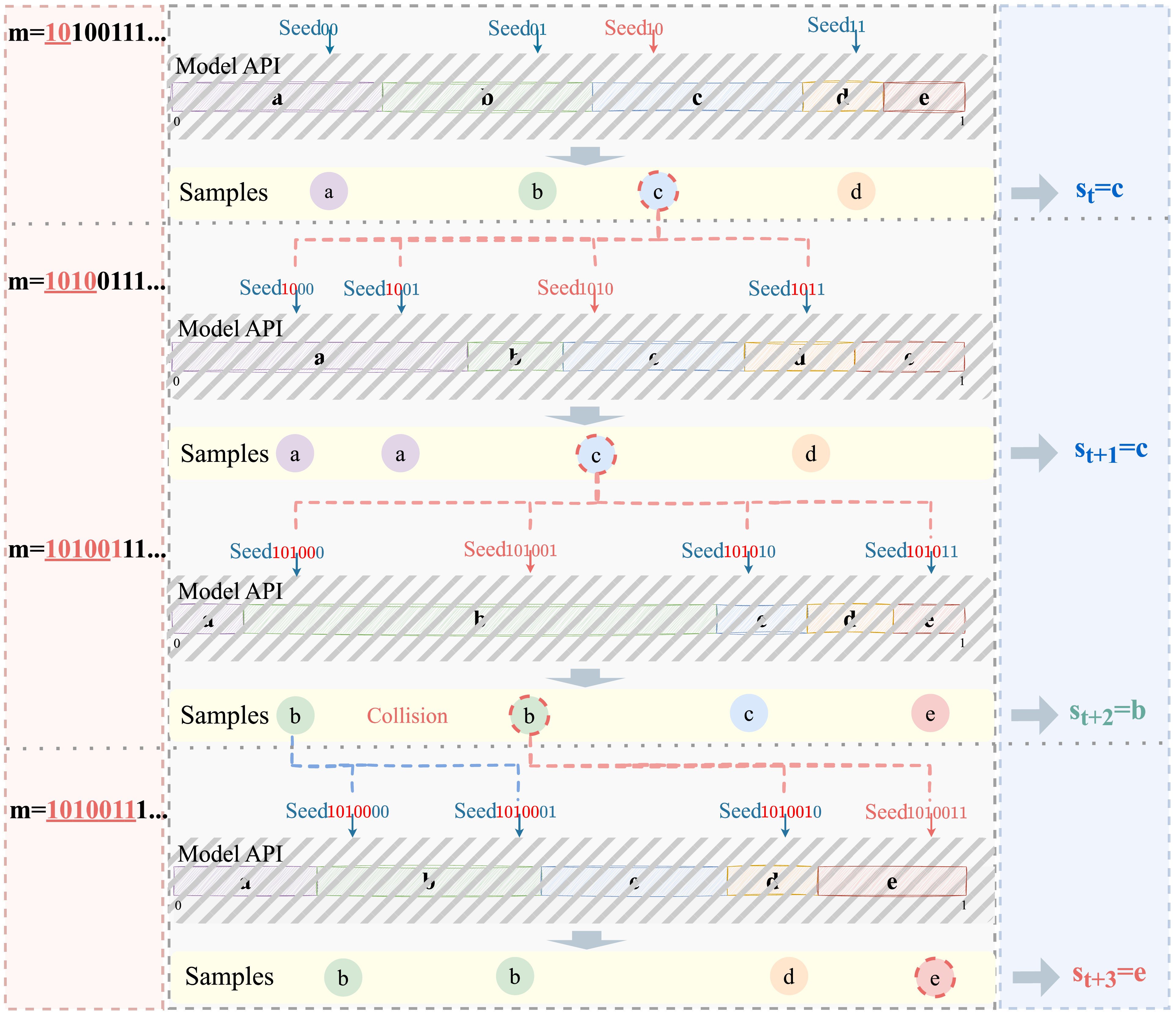}
    \caption{An example of encoding algorithm~\ref{encoding_algo} when the maximum size of the collision set is set to 4 at each time step ($N$=2). Here, $m$ represents the secret message, $s_t$ is the stego token output at time $t$, and $seed_{10}$ represents the seed for message $10$. The arrows ($\downarrow$) indicate the sampling outcomes (balls) based on the $seed$ determined by the candidate messages. \textcolor{red}{Red secret message in $\textbf{m}$} denotes the already embedded secret bits, and \textcolor{red}{\underline{underlined text in $\textbf{m}$}} represents the secret message that can be immediately extracted.}
    \label{fig:example}
\end{figure*}

Here we have 2 key insights to design our steganography scheme. The first insight is about the sampling strategy. If we sample multiple candidate tokens from the API and uniformly choose one of them to output, the distribution of the output token should be the same as the channel distribution. Suppose that we have taken $2^N$ samples, we can construct a one-to-one map from all of the $N$-bit strings to these samples. Obviously, this map is a function from $\{0,1\}^N$ to the model's vocabulary $V$, denoted as $F: \{0,1\}^N \xrightarrow{} V$. Because the messages exist before we query for these samples, we can use the pseudorandom generator to generate a seed $seed_{m}$ for each message $m$. And then use the seed to sample from the model API to construct the map $F(m) := \mathsf{Model}_{seed_m}(h)$.  As Figure \ref{fig:example} shows, if we sample four times, the four samples can correspond to fragments of the possible secret messages: "00", "01", "10", and "11" respectively. If the secret message starts with "10", then we choose the token corresponding to "10" to output. If the secret message is randomized by XOR, then the output token should obey the distribution of the model.
 
In some lucky cases when the secret message corresponds uniquely to a sampling result, we can decode the secret message at that moment. However, if the sampling result corresponding to one possible secret message is the same as that of another possible secret message, this is known as a \textit{collision} (time $t+2$ in Figure \ref{fig:example}, where the "101001" and "101000" are mapped to the same sample). In METEOR, the collision forces the decoding algorithm to extract the shared prefix of all collision strings. But we think there is a patient and more effective strategy. Our second insight is about the collision phenomenon. 

It is easy to observe that a collision does not affect the embedding process of the secret message. It only affects the decoding process. So when the collision happens, 
We do not avoid the collisions but record all of the strings colliding with the true secret message in a list (including the secret message). In the next encoding step, in order to construct the one-to-one map $F$, the number of samples should be the same as the number of elements in the recorded collision set. Within a few encoding steps, the size of the collision set should shrink to $1$, which only contains the secret message. 

However, decreasing the number of samples may bring in some unlucky cases. For example, when there are only 2 elements in the collision set, one of them is the true message while the other is a fake one, we may continuously get 2 identical samples and take a long time to distinguish the true message. Moreover, if the model API outputs an end-of-text token before the above process completes, we can only extract the shared prefix of them. Such a problem has a negative effect on the capacity of the steganography scheme. Therefore, it is better to maintain the size of the collision set at a high level, to ensure that in each step we can rule out a part of the messages. This observation requires that we expand the collision set in a reasonable way. We know that for an $N$-bit prefix of the secret message, the next $N+1$ bit is either 1 or 0. So we can append $1$ and $0$ to all of the elements in the collision set to double its size. To avoid the out-of-memory bugs when running our algorithm, we often set a limit for the size of the collision set. When we set the maximum size of the list to $2^n$, in the encoding procedure its size will be bounded by $[2^{n-1}, 2^n]$. This is because once its size is smaller than $2^{n-1}$, it will trigger the expansion operation to double its size, until its size is larger than $2^{n-1}$. This expansion operation ensures that the number of samples is not too small, which ensures the high capacity of our scheme. At the end of the text, we can extract the shared prefix of the collision strings. The whole procedure avoids the capacity loss induced by frequently extracting the shared prefix as METEOR does.

\subsection{Our Steganography Scheme}

In this subsection, we formally describe the encoding and decoding algorithms of our steganography scheme $\Sigma_{\mathsf{Model}} = (\mathsf{KeyGen}, \mathsf{Enc}_{\mathsf{Model}, k, N}, \mathsf{Dec}_{\mathsf{Model}, k, N})$. Here we introduce a new parameter $N$ to control that the maximum size of collision set is less than $2^N$. 

\subsubsection{Encoding (Adaptive Dynamic Sampling)}

The encoding algorithm can be divided into two steps: determining the number of samples, sampling, and identifying collision samples. The pseudocode for the overall encoding algorithm is presented in Algorithm \ref{alg:encoding-all}.
\begin{figure*}[htb]
	\centering
	\includegraphics[width=\linewidth]{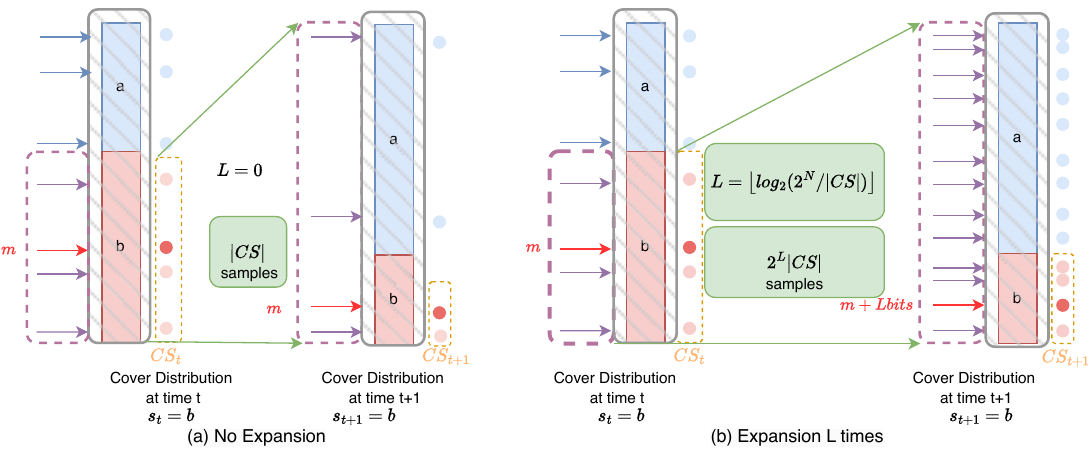}
	\caption{Illustration of dynamically determining the number of samples and expanding the secret. (a) Non-expansion case $L=0$: at step $t+1$, the number of samples $|CS|$ is the size of the conflict set at step $t$. (b) General expansion case $0<L\le N$: at step $t+1$, the number of samples is $2^{L}\!\cdot\!|CS|$.}
	\label{expapnsion}
\end{figure*}

In the initial step, the encoding algorithm sets the collision set $CS$ as $\{0,1\}^N$, which includes the $N$-bit prefix of the secret message.  At each step, we have to construct a one-to-one map $F$ to relate the possible message to each sample. For every element $e$ in the collision set $CS$, the encoding algorithm independently queries the model API for a sample $t$, and let $F(e) := t$. For each query, it first uses a pseudorandom generator with a key $k$ to generate the seeds for the model API and then invokes the model API with the seed and history to get the output token $t$. After the construction of map $F$, the encoding algorithm uses the sample $s_i = F(m_{1:l})$ (corresponding to the prefix of secret message $m_{1:l}$) as the next token of stegotext. Then the encoding algorithm filters the elements of the collision set whose image under $F$ is the same as $s_i$ to construct the updated collision set, which can be denoted as $\{e \in CS \mid F(e) = s_i\}$. After updating the collision set, it will be expanded if possible. If $|CS|$ is less than $2^{N-1}$, we expand the collision set by appending $0$ and $1$ to every element in the list, which ensures that the size of the collision set is between $2^{N-1}$ and $2^N$. If $|CS| > 2^{N-1}$, then we do not expand the collision set, and the next step embeds no new secret bits and simply draws $|CS|$ samples to rule out more fake messages. If the current collision set size $|CS|$ is small, we may expand the list multiple times in the next step and thereby introduce more secret bits. For example, the $2$-bit prefix of secret message $10_2$ (i.e., 2 in decimal) after 2 times of expansion will be four candidates: $1000_2\ (8),\ 1001_2\ (9),\ 1010_2\ (10),\ 1011_2\ (11)$. One of them must be the $4$-bit prefix of the secret message.  After $L = \lfloor 2^N / |CS|\rfloor$ times of expansion, the length of the encoded secret message will increase by $L$ bits, and the expanded collision set is carried into the next step. Figure~\ref{expapnsion} illustrates the dynamic expansion mechanism for the cases $L=0$ (no expansion) and $0<L\le N$ (expand $L$ times).


In other words, the encoding procedure can be viewed as the flow control.
If collision occurs frequently, the window size decreases additively; otherwise, it increases multiplicatively. 

\begin{algorithm}[ht]

\caption{$\mathsf{Enc}_{\mathsf{Model},k,N}(m,h) \xrightarrow{} s$}
    \renewcommand{\algorithmicrequire}{\textbf{Input:}}
    \renewcommand{\algorithmicensure}{\textbf{Output:}}
    \label{alg:encoding-all}
    \begin{algorithmic}[1]
    \label{encoding_algo}
    \REQUIRE Language model API $\mathsf{Model}_{seed}(\cdot)$, history $h$, key $k$, secret message $m$, maximum size of collision set $2^N$, pseudorandom generator $\mathsf{PRG}_k(\cdot)$
    \ENSURE Stegotext $s$.
    \STATE $s\longleftarrow{\emptyset}$
    \STATE $i,l \longleftarrow 0,N$
    \STATE $CS_i \longleftarrow\{0,1\}^N$ $//$ \textit{Initialize the collision set $CS$.}
    \WHILE{ not the end of text}
       \FOR{element $e \in CS_i$}
           \STATE $seed_e \longleftarrow \mathsf{PRG}_k(\cdot)$
           \STATE $t \longleftarrow \mathsf{Model}_{seed_e}(h)$
           \STATE Let $F(e) := t$ $//$ \textit{Construct the message-to-token map $F$.}
       \ENDFOR
       \STATE $s_i \longleftarrow F(m_{1:l})$ $//$ \textit{Choose the token corresponding to the l-bit prefix of secret message.}
       \STATE $s,h \longleftarrow s || s_i, h || s_i$ $//$ \textit{Append the new token $s_i$ to stegotext $s$ and history $h$ for the next step.}
       \STATE $CS_{i+1} \longleftarrow \{e \in CS_i \mid F(e) = s_i\}$ $//$ \textit{Update collision set.}
       \WHILE{$|CS_{i+1}| < 2^{N-1}$}
           \STATE $CS_{i+1} \longleftarrow \{e || b \mid e \in CS_{i + 1}, b\in\{0,1\}\}$ $//$ \textit{Expand collision set.}
           \STATE $l \longleftarrow l+1$
       \ENDWHILE
       \STATE $i \longleftarrow i+1$
       \ENDWHILE 
\RETURN $s$
    \end{algorithmic}
\end{algorithm}

   


\begin{algorithm}[ht]
\caption{$\mathsf{Dec}_{\mathsf{Model},k,N}(s,h) \xrightarrow{} m$ }
    \renewcommand{\algorithmicrequire}{\textbf{Input:}}
    \renewcommand{\algorithmicensure}{\textbf{Output:}}
    \label{alg:decoding-all}
    \begin{algorithmic}[1]
    \REQUIRE Stegotext $s$, language model API $\mathsf{Model}_{seed}(\cdot)$, history $h$, key $k$, maximum size of collision set $2^N$, pseudorandom generator $\mathsf{PRG}_k(\cdot)$
    \ENSURE Secret message $m$.
    \STATE $i \longleftarrow 0$
    \STATE $CS_i\longleftarrow\{0,1\}^N$
    \FOR{ $i \in {1,\cdots,|s|}$ }
        \FOR{element $e \in CS_i$}
           \STATE $seed_e \longleftarrow \mathsf{PRG}_k(\cdot)$
           \STATE $t \longleftarrow \mathsf{Model}_{seed_e}(h)$
           \STATE Let $F(e) := t$ $//$ \textit{Construct the one-to-one map $F$.}
       \ENDFOR
       \STATE $h \longleftarrow h || s_i$
       \STATE $CS_{i+1} \longleftarrow \{e \in CS_i \mid F(e) = s_i\}$ $//$ \textit{Update collision set.}
       \WHILE{$|CS_{i+1}| < 2^{N-1}$}
           \STATE $CS_{i+1} \longleftarrow \{e || b \mid e \in CS_{i + 1}, b\in\{0,1\}\}$ $//$ \textit{Expand collision set.}
       \ENDWHILE
       \STATE $i \longleftarrow i+1$
    \ENDFOR
\RETURN $m \longleftarrow $ shared prefix of all elements in $CS_i$
    \end{algorithmic}
\end{algorithm}

\subsubsection{Decoding}
The decoding algorithm is almost the same as the encoding algorithm,  as shown in Algorithm \ref{alg:decoding-all}. As the key is shared, the decoding algorithm can generate the same seeds and sample from the model API with them. Therefore, the decoding algorithm can get the same samples as the encoding algorithm, which ensures that the one-to-one map $F$ constructed by the decoder is the same as the encoder. Then the decoding algorithm can identify the collision samples and update the collision set. After that, if the size of the collision set is smaller than $2^{N-1}$, it will be expanded until its size is greater than $2^{N-1}$. This process continues until all tokens in the stegotext are processed, then we can extract the shared prefix of the collision set, which is the encoded secret message $m$.

\subsection{Proof of security}
Firstly, we consider the case where only 1 bit of the secret message  ($m_i=0$ or $1$) is embedded. $P[m_i=0]+P[m_i=1]=1.$ Then we sample twice from the model distribution can get $2$ sample results, namely $c_0$ and $c_1$.
For any symbol (token) $a$ in the model’s vocabulary, let $\Pr[\mathsf{Model}(h)\xrightarrow{}a]$ denote the probability that the model $\mathsf{Model}$ would output the token $a$ under its standard generation process. Let $\Pr[\mathsf{Enc}_{\mathsf{Model},k,N}(m_i,h) \xrightarrow{}a]$ denote the probability that the encoding algorithm outputs token $a$, which is given by:
\begin{align*}
    & \Pr[\mathsf{Enc}_{\mathsf{Model},k,N}(m_i,h) \xrightarrow{}a]\\
    =& \Pr[m_i=0]\cdot \Pr[\mathsf{Enc}_{\mathsf{Model},k,N}(0,h) \xrightarrow{}a] \\
    &+\Pr[m_i=1]\cdot \Pr[\mathsf{Enc}_{\mathsf{Model},k,N}(1,h) \xrightarrow{}a]  \\
     =& \Pr[m_i=0]\cdot \Pr[c_0=a] + \Pr[m_i=1]\cdot \Pr[c_1=a] \\
             =& \Pr[m_i=0]\cdot \Pr[\mathsf{Model}(h)\xrightarrow{}a] 
             \\&+ \Pr[m_i=1]\cdot \Pr[\mathsf{Model}(h)\xrightarrow{}a] \\
             =& \Pr[\mathsf{Model}(h)\xrightarrow{}a].
\end{align*}

Similarly, this conclusion also holds when embedding multiple bits of information because our steganography schemes only sample multiple times from the model and do not distort the original model distribution. 

From a more formal standpoint, we prove the computational-indistinguishable security by a sequence of games $G_0 \sim G_3$, as shown in Figure \ref{games}. The security of our framework relies on the pseudo-random generator and the i.i.d. property of samples drawn from the same distribution.

$G_0 = G_1$: $G_0$ is the steganographic sampling procedure controlled by the secret message $m$, while $G_1$ is independent of secret message. The probability that $G_1$ outputs the token $a$ is
{
\begin{align*}
    \Pr[G_1 \xrightarrow{} a] &= \sum_{e \in CS}\frac{1}{|CS|}\cdot \Pr[\mathsf{Model}_{\mathsf{PRG}_k(\cdot)}(h) \xrightarrow{} a] \\
             &=\Pr[\mathsf{Model}_{\mathsf{PRG}_k(\cdot)}(h) \xrightarrow{} a]
\end{align*}
}
due to the independent sampling procedure, this probability is the same as $\Pr[G_0 \xrightarrow{} a]$.

$G_1 \approx_c G_2$: $G_2$ uses random seeds to substitute the pseudo-random generator $\mathsf{PRG}_k(\cdot)$ in $G_1$. From the definition of pseudo-random generator, for any probabilistic polynomial time (p.p.t.) adversary $\mathcal{A}$, 
{
\begin{align}
    \left|\Pr\left[ \mathcal{A}^{\mathsf{PRG}_k(\cdot)}(1^\lambda) = 1\right] - \Pr\left[\mathcal{A}^{R(\cdot)}(1^\lambda) = 1\right]\right| \leq \mathsf{negl}(\lambda),
\end{align}
}
where the $\lambda$ is the security parameter of $\mathsf{PRG}$, $1^\lambda$ denotes the internal randomness of adversary $\mathcal{A}$.
Assuming the existence of a p.p.t. adversary $\mathcal{A}'$, which can distinguish the output of $G_1$ and $G_2$ with a noticeable probability, we can use the adversary $\mathcal{A}'$ to construct another p.p.t. adversary $\mathcal{A}''$ to distinguish between the output of the random generator $R(\cdot)$ and pseudo-random generator $\mathsf{PRG}_k(\cdot)$. $\mathcal{A}''$ runs $G_1$ and $G_2$ with seeds generated by random generator $R(\cdot)$ and pseudo-random generator $\mathsf{PRG}_k(\cdot)$, respectively. Though now  $\mathcal{A}''$ can neither distinguish between $R(\cdot)$ and $\mathsf{PRG}_k(\cdot)$ nor $G_1$ and $G_2$, he can query $\mathcal{A}'$ for each pair of output of $G_1$ and $G_2$. Since $\mathcal{A}'$ can distinguish the output of $G_1$ and $G_2$ with a noticeable probability, the adversary $\mathcal{A}''$ can distinguish the output of the random generator $R(\cdot)$ and pseudo-random generator $\mathsf{PRG}_k(\cdot)$ with a noticeable probability, which contradicts the definition of pseudo-random generator. Therefore, such adversary $\mathcal{A}'$ does not exist, and the distributions of output of $G_1$ and $G_2$ are computationally indistinguishable.

$G_2 = G_3: G_2$ leverages the independent and identically distributed (i.i.d.) property of samples drawn from the same distribution. Since the output of the sampling process follows the model's distribution, randomly sampling a result from multiple samples is equivalent to directly sampling a result from the model itself.

Up to this point, we have demonstrated the security of our steganography scheme.

\begin{figure}[htbp]
    \centering
    \begin{theobox}{$G_0$: \textsf{Steganography-Sampler}}
    \begin{algorithmic}[1]
        \FOR{$e \in CS$}
            \STATE $seed \longleftarrow \mathsf{PRG}_k(\cdot)$
            \STATE $t \longleftarrow \mathsf{Model}_{seed}(h)$
            \STATE Let $F(e) := t$
        \ENDFOR
        \STATE $s_i \longleftarrow F(m_{1:l})$
        \RETURN $s_i$
    \end{algorithmic}
    \end{theobox}
    \begin{theobox}{$G_1$:  \textsf{Pseudorandom Multiple Sampling }}
    \begin{algorithmic}[1]
        \FOR{$e \in CS$}
            \STATE $seed \longleftarrow \mathsf{PRG}_k(\cdot)$
            \STATE $t \longleftarrow \mathsf{Model}_{seed}(h)$
            \STATE Let $F(e) := t$
        \ENDFOR
        \STATE Randomly choose $e^* \in CS$
        \STATE $s_i \longleftarrow F(e^*)$
        \RETURN $s_i$
    \end{algorithmic}
    \end{theobox}
    \begin{theobox}{$G_2$:  \textsf{Random Multiple Sampling}}
    \begin{algorithmic}[1]
        \FOR{$e \in CS$}
            \STATE Randomly generate $seed$
            \STATE $t \longleftarrow \mathsf{Model}_{seed}(h)$
            \STATE Let $F(e) := t$
        \ENDFOR
        \STATE Randomly choose $e^* \in CS$
        \STATE $s_i \longleftarrow F(e^*)$
        \RETURN $s_i$
    \end{algorithmic}
    \end{theobox}
    \begin{theobox}{$G_3$:  \textsf{Model Sampling}}
    \begin{algorithmic}[1]
            \STATE Randomly generate $seed$
            \STATE $t \longleftarrow \mathsf{Model}_{seed}(h)$
    \end{algorithmic}
    \end{theobox}
    \caption{Games used in the proof of steganography security.}
    \label{games}
\end{figure}

\section{Experiment \& Discussion}

\subsection{Experiment Settings}

\textbf{Task}
Theoretically, our steganography method can be applied to a wide range of carriers, including text, audio, and images. In the following discussion, We primarily use text generation tasks to demonstrate out scheme’s effectiveness, as text generation with large language models is currently widely popular.

\textbf{Baseline}
We evaluated our adaptive dynamic sampling steganography method using the random sampling decoding approach, alongside the ideal steganography algorithm and provably secure generative steganography algorithms, including METEOR \cite{meteor2021}, DISCOP \cite{ding2023discop}, its capacity-boosted variant DISCOP Reorder, SparSamp~\cite{wang2025sparsamp}, and iMEC~\cite{IMECde2022perfectly}. Each baseline is evaluated using its default parameter configuration. DISCOP and its variants were implemented in Cython, while the remaining methods were implemented in Python. 

\textbf{Dataset}
We used three types of datasets to simulate cover text channels. The first is a uniform distribution, simulating high entropy situations where each token in the language model’s tokenizer vocabulary has an equal probability. The second is a QA dataset, using the WILD \cite{wild}, a typical application scenario where the language model answers questions. Finally, we used the XHS \cite{xhs} dataset, which simulates social media text in Chinese, characterized by an informal, fragmented style typical of social network communication.

\textbf{Model}
We evaluate our steganography method on three large language models: \textsc{Mistral3} \cite{mistral}, \textsc{Qwen2} \cite{qwen}, and \textsc{Llama3} \cite{llama3}, using the WILD and XHS datasets, each with a model scale of roughly 7B parameters. We customize the vocabulary size for testing under the uniform distribution.
The prompts used for WILD and XHS datasets correspond to the questions within each dataset. These prompts are loaded into the function {\texttt{apply\_chat\_template}} and then become the input of LLMs.
For each test, we let the models generate 1,000 samples with  top-$p = 1.0$, base $temperature =1.0$, and the max generated length is 500.
All our experiments were conducted on 2 $\times$ NVIDIA A5000 GPUs (32GB RAM) and 24 $\times$ Intel Xeon w5-3423 CPUs.

\subsection{Metrics}

We evaluate the performance of our steganography scheme in terms of \textbf{Security, Efficiency, Capacity, and Correctness}.

\subsubsection{Security}
Although our steganography scheme's computational security has been theoretically proven in the Method section, we further validate its security through linguistic steganalysis in adversarial scenarios and assess its intuitive imperceptibility based on linguistic quality.

\textbf{Steganalysis}. We utilized a classic steganalysis model to assess the imperceptibility of stegotexts: TS-CSW \cite{TS-CSW} based on pre-trained BERT \cite{BERT}.
We trained the detectors on 1,000 samples of generated stegotext and corresponding cover texts, randomly sampled from the same language model, using a 3:1:1 split for training, validation, and testing. With a learning rate of 1e-4, training was conducted for three epochs. After repeating the process three times, the average F1 score from the test set was used as the steganalysis F1 score.

\textbf{Linguistic Quality.} We evaluate linguistic quality using perplexity (PPL) and diversity. PPL measures the fluency of generated text, with lower PPL values indicating smoother and more natural text.
\begin{equation}
    PPL = \exp\left(-\frac{1}{W}\sum_{i=1}^{W}\log \ \Pr[x_i|{\bf x}_{1 : i - 1 }]\right),
\end{equation}
where $W$ is the length of the generated text.

As for text diversity, we used the $ distinct_n$ metric. This metric needs to find the unique pieces of tokens in the text and calculate their ratio. 
\begin{equation}
    distinct_n = \frac{\text{count(unique\, n-grams)}}{\text{count(n-grams)}}.
\end{equation}

\subsubsection{Efficiency}
\textbf{Time.}
For real-time covert communication, it’s important to complete stego embedding quickly. To measure this efficiency, we measure the average time of stego generation, calculated by dividing the total time taken for the entire process by the length of the embedded message. The shorter, the better.

\subsubsection{Capacity}

\textbf{Entropy.} This represents the theoretical upper bound of the embedding capacity, measured in bits per symbol. For texts, a symbol usually refers to a token generated by large language models. Entropy is calculated as: 
\begin{equation}
Entropy = \frac{1}{W}\sum_i^W \Pr[x_i|x_{<i}]\log_2\Pr[x_i|x_{<i}].
\end{equation}
\textbf{Embedding Capacity.} This metric measures the average number of bits that can be successfully embedded and extracted, expressed in bits per token. 

\textbf{Utilization.} We quantify how efficiently the scheme exploits generative entropy by
$U \;=\; \frac{C_{\text{eff}}}{H} \in [0,1],$
where $C_{\text{eff}}$ is the reliably decodable embedding rate (bits/token) and $H$ is the per-token Shannon entropy of the model. Larger $U$ indicates better entropy utilization; $U=1$ corresponds to achieving the theoretical limit.

\subsubsection{Correctness}
\textbf{Success Rate}
Following iMEC~\cite{IMECde2022perfectly}, we evaluate correctness using the bit-level decoding Success Rate (SR).
Let $\mathbf{b}$ be the original bit sequence and $\hat{\mathbf{b}}$ the decoded bit sequence, both of length$ |\mathbf{b}|$. We define
\begin{equation}
SR = \frac{1}{|\mathbf{b}|}\sum_{j=1}^{|\mathbf{b}|} \mathbf{1}[b_j = \hat{b}_j],
\end{equation}
where $\mathbf{1}[\cdot]$ is the indicator function. A value of 1 means perfectly correct decoding, while smaller values correspond to higher bit error rates.

\subsection{Evaluation Results and Discussion}

\begin{table*}[ht]
    \centering
     \resizebox{\textwidth}{!}{{
     \color{black}
    \begin{tabular}{cccccccccccc}
    \toprule[1.5pt]
     \multirow{2}{*}{\textbf{Model}}&\multirow{2}{*}{\textbf{Algorithm}}&  \multicolumn{3}{c}{\textbf{Capacity}}& \multicolumn{1}{c}{\textbf{Time}} & \multicolumn{4}{c}{\textbf{Linguistic Quality}}&\textbf{Imperceptibility}&\textbf{Correctness}\\
         \cmidrule(lr){3-5}
         \cmidrule(lr){6-6}
        \cmidrule(lr){7-10}
                \cmidrule(lr){11-11}
                  \cmidrule(lr){12-12}
         &  &  \makecell{Entropy \\ bit/token}&  \makecell{Embed \\ bit/token}$\uparrow$ &  Utili.$\uparrow$ &  \makecell{Time\\sec./token}$\downarrow$ &  PPL &  Dist2 &  Dist3 & Dist4 & TS-CSW F1$\downarrow$&SR$\uparrow$\\\midrule
 \multirow{7}{*}{\textsc{Mistral}}& Random Sampling& 0.9489
& -
& -
&  0.0295&  1.9890
&  0.6073
&  0.8377
&  0.9208& -&-
\\
 & METEOR &   0.9870
&   0.5732& 0 .5633&  0.0298&  2.0888
&  0.6135
&  0.8458
&  0.9275& $54.23\%_{3.95}$&1.0000
\\
 & DISCOP (R.) & 0.9506
& 0.7683& 0.7930&0.0427& 2.0141
& 0.6086
& 0.8389
& 0.9215& $51.42\%_{1.24}$&1.0000
\\
 & DISCOP & 0.9267
& 0.4242& 0.4631& 0.0379& 1.9848
&  0.6044
& 0.8351
&  0.9192&$51.51\%_{1.32}$&1.0000\\
 & SparSamp & 1.0144
& 0.8562& 0.8411& 0.0314& 2.1201
&  0.6091
& 0.8426
&  0.9245&$51.57\%_{1.16}$&1.0000\\
 & iMEC & 1.0235
& 0.7345& 0.6923&3.8231& 2.2813
&  0.6865
& 0.8827
&  0.9442&$53.25\%_{1.70}$&0.8825
\\\cmidrule{2-12}
 & Ours&0.9872&\textbf{0.8964}& \textbf{0.9137}& 0.0295& 2.0788
& 0.6097
& 0.8415
& 0.9246&$50.18\%_{1.62}$&1.0000
\\
\midrule
 \multirow{7}{*}{\textsc{Qwen}}& RandomSampling& 2.0806
& -
& -
& 0.0296& 4.7768
& 0.6604
&0.8750
&  0.9382  &  -    &-                                                        
\\
 & METEOR& 2.0963
&1.3047& 0.6233& 0.0329& 4.9355
& 0.6676
& 0.8788
& 0.9426 & $51.83\%_{2.19}$&1.0000
\\
 & DISCOP(R.)&  2.1539
&1.9093&0.8866& 0.0822&  5.1384
&0.6708
& 0.8821
& 0.9431 & $51.48\%_{2.19}$&1.0000
\\
 & DISCOP&1.9569
&0.8367&0.4275& 0.0558&4.2501
& 0.6632
&0.8748
&0.9394& $51.94\%_{1.27}$&1.0000
\\
 & SparSamp&2.0895
&1.8905&0.9052& 0.0299&4.9285
& 0.6431
&0.8919
&0.9426 &$51.87\%_{1.36}$&1.0000\\
&iMEC&2.1231&1.6174&0.7466
& 3.5392
&4.9894
&0.6994
&0.9023 &0.9536&$52.27\%_{3.19}$&0.9952
\\\cmidrule{2-12}
 & Ours& 2.0641&1.8789& \textbf{0.9103}&\textbf{0.0293}& 4.7569
&0.6716
&0.8811
&0.9409&$49.83\%_{1.58}$&1.0000
\\
\midrule
 \multirow{7}{*}{\textsc{Llama3}}& RandomSampling&  0.6934
& -
& -
&0.0312& 1.7340
& 0.6090
& 0.8399
& 0.9222 &-&-

\\
 & METEOR&0.7172
& 0.4031&0.5574& 0.0317& 1.7815
& 0.6100
& 0.8465
& 0.9284 &$53.86\%_{2.39}$&1.0000
\\
 & DISCOP(R.)& 0.6955
& 0.5221& 0.7425& 0.0765& 1.7083
& 0.6060
& 0.8408
& 0.9219 & $50.19\%_{1.21}$&1.0000
\\
 & DISCOP& 0.6928
& 0.3297& 0.4795& 0.0540& 1.7405
&0.6078
& 0.8456
& 0.9301 &$49.98\%_{1.57}$&1.0000\\
 & SparSamp&0.7085
&0.5713&0.8183&0.0336&1.7657
& 0.6061
&0.8440
&0.9293 &$51.07\%_{1.31}$&1.0000\\
 & iMEC&0.8390
&0.6057&0.7218& 5.9045&1.8825
& 0.6494
&0.8822
&0.9416 &$55.17\%_{2.27}$&0.8432
\\\cmidrule{2-12}
 & Ours&0.7028& \textbf{0.6572}& \textbf{0.9352}&0.0313& 1.7035
& 0.6063
& 0.8339
& 0.9260 &$49.73\%_{2.79}$&1.0000
\\\bottomrule[1.5pt]
         
    \end{tabular} }}
    \caption{Main Results on WILD. Higher capacity and utilization are better. Lower time is better. Linguistic quality closer to random sampling is better. In our method, the maximum size of collision set $2^N$ can be expanded at each time is set to $2^9$.}
    \label{tab:wild}
\end{table*}

\begin{table*}[ht]
    \centering
     \resizebox{\textwidth}{!}{{
     \color{black}
    \begin{tabular}{cccccccccccc}
    \toprule[1.5pt]
        \multirow{2}{*}{\textbf{Model}}&\multirow{2}{*}{\textbf{Algorithm}}&  \multicolumn{3}{c}{\textbf{Capacity}}& \multicolumn{1}{c}{\textbf{Time}} & \multicolumn{4}{c}{\textbf{Linguistic Quality}}&\textbf{Imperceptibility}&\textbf{Correctness}\\
         \cmidrule(lr){3-5}
         \cmidrule(lr){6-6}
        \cmidrule(lr){7-10}
                \cmidrule(lr){11-11}
                   \cmidrule(lr){12-12}
         &  &  \makecell{Entropy \\ bit/token}&  \makecell{Embed \\ bit/token}$\uparrow$ &  Utili.$\uparrow$ &  \makecell{Time\\sec./token}$\downarrow$ &  PPL &  Dist2 &  Dist3 & Dist4 & TS-CSW F1$\downarrow$&SR$\uparrow$\\\midrule
       \multirow{7}{*}{ \textsc{Mistral}}&  RandomSampling
&  2.1963
&  -
&  -
&   0.0294& 4.8066
&  0.5367
&  0.7639
& 0.8657 & - &-
\\
         &  METEOR 
&  2.2195
& 1.5028&  0.6635&  0.0299&  4.8673
&   0.5183
& 0.7560
&0.8614& $52.91\%_{3.83}$&1.0000
\\
         &  DISCOP(R.)
&  2.2863
&  2.0416&   0.8825& 0.0427&  5.0154
&  0.5345
&  0.7723
& 0.8792&$51.84\%_{3.67}$&1.0000
\\
         &  DISCOP&  2.2291
& 0.8450& 0.3837& 0.0370& 4.8425
& 0.5219
& 0.7564
& 0.8620&$50.97\%_{3.03}$&1.0000\\
 & SparSamp&2.3017
&2.0893&0.9029&0.0313&5.0892
& 0.5458
&0.7869
&0.8824 &$51.83\%_{2.17}$&1.0000\\\
 & iMEC&2.4382 
&2.0871&0.8579& 5.1984&5.1785
& 0.5655
&0.8139
&0.8947 &$53.29\%_{2.33}$&0.9847
\\\cmidrule{2-12}
         &  Ours& 2.1986& 2.0378& \textbf{0.9308}& 0.0298& 4.9094&  0.5341&  0.7675 & 0.8721&$50.81\%_{2.12}$&1.0000\\
 \midrule
               \multirow{7}{*}{   \textsc{Qwen}}&  RandomSampling
& 3.5673
& -
& -
& 0.0319&  12.6965
&  0.7527
& 0.9198
& 0.9585 &-&-
\\
 & METEOR& 3.6247
& 2.5283 &0.6892&0.0328& 13.4175
& 0.7634
& 0.9309
&0.9704 & $52.38\%_{3.10}$&1.0000
\\
 & DISCOP(R.)& 3.5478
&3.2606& 0.9118&0.0824& 14.7447
& 0.7498
& 0.9171
&0.9542 &$50.28\%_{2.64}$&1.0000
\\
 & DISCOP& 3.5371
& 1.3595&0.3852&0.0523& 12.8203
& 0.7365
& 0.9066
&0.9467 & $50.03\%_{2.96}$&1.0000\\
 & SparSamp&3.6482
&3.4095&0.9304& 0.0334&13.5791
& 0.7549
&0.9241
&0.9771 &$51.29\%_{1.87}$&1.0000\\
 & iMEC&3.1611
&2.6495&0.8416& 4.2276&13.1721
& 0.7638
&0.9088
&0.9665 &$53.52\%_{1.21}$&0.9987
\\\cmidrule{2-12}
         &  Ours&3.7272& \textbf{3.4928}&  \textbf{0.9393}&\textbf{0.0299} &13.3314& 0.7642& 0.9209& 0.9588&$49.83\%_{2.58}$&1.0000\\
          \midrule
 \multirow{7}{*}{\textsc{llama3}}& RandomSampling& 1.3315
& -
& -
& 0.0313& 3.3848
& 0.5394
& 0.7649
&0.8635& -&-
\\
          &  METEOR
&  1.3410
& 0.8466&  0.5760&  0.0319&  7.4924
&  0.5427
& 0.7720
&0.8743 & $52.38\%_{3.10}$&1.0000
\\
         &  DISCOP(R.)
&  1.4737
&  1.2137&  0.8243& 0.0762&  4.9633
&  0.5583
&  0.7762
& 0.8721 &$50.01\%_{2.70}$&1.0000
\\
         &  DISCOP&  1.3739
& 0.5625&  0.4380& 0.0519& 4.3652
& 0.5399
& 0.7639
& 0.8678 &$49.98\%_{2.09}$&1.0000\\
 & SparSamp&1.2692
&1.4206&0.9105& 0.0343&4.4513
& 0.5582
&0.7857
&0.8874 &$50.73\%_{2.17}$&1.0000\\
 & iMEC&1.6817
&1.3383&0.8177&4.9624&4.1399
& 0.5976
&0.8132
&0.9301 &$54.24\%_{3.19}$&0.8631
\\\cmidrule{2-12}
   & Ours& 1.4267& 1.2560&\textbf{ 0.9223}& 0.0314&4.3966& 0.5447& 0.7727&0.8745&$49.66\%_{2.58}$&1.0000\\\bottomrule[1.5pt]
    \end{tabular}
    }}
    \caption{Main Results on XHS. Higher capacity and utilization are better. Lower time is better. Linguistic quality closer to random sampling is better. In our method, the maximum size of the collision set $2^N$ can be expanded at each time is set to $2^9$.}
    \label{tab:XHS}
\end{table*}


\begin{figure*}
    \centering
    \includegraphics[width=1\linewidth]{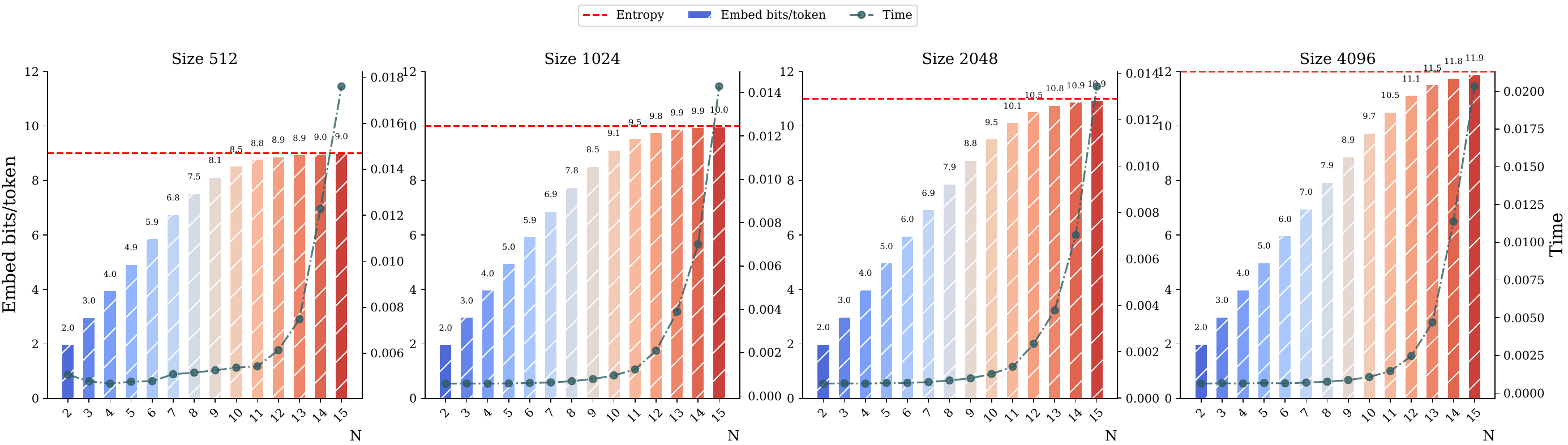}
    \caption{The relationship between the number of uniform distribution samples and embedding capacity. $N$ represents the maximum number of samples as $2^{N}$, the red dashed line indicates entropy, and the green line represents the generation time per token.}
    \label{fig:uniform}
\end{figure*}

\subsubsection{On the Capacity \& Entropy}

As shown in Tables \ref{tab:wild} and \ref{tab:XHS}, our method achieves the highest embedding capacity and entropy utilization, even outperforming DISCOP Reorder and SparSamp, which have the best capacity under explicit distribution.
Our work was developed contemporaneously with SparSamp~\cite{wang2025sparsamp}, a provably secure steganography scheme, but our ADS method differs in a fundamental way. SparSamp, building on DISCOP~\cite{ding2023discop}, performs sparse sampling for \textit{fixed-length} secret blocks and only proceeds to the next block once the current block can be decoded unambiguously. Beyond implementation details, our approach is conceptually distinct: ADS \textit{expands} the secret adaptively during generation, thereby exploiting a larger fraction of the text’s entropy and achieving higher embedding capacity.

The results from the uniform distribution in Figure~\ref{fig:uniform} show that as the maximum allowed sampling approaches the vocabulary size, the embedding capacity asymptotically approaches the entropy of the uniform distribution. More sampling iterations result in higher embedding capacity. This trend holds across all models with different vocabulary sizes. For all models and scenarios, our method achieves high entropy utilization, even when the distribution is unknown. However, the upper limit of embedding capacity still depends on the distribution’s entropy.

\subsubsection{Time Efficiency}

The runtime of our method is primarily constrained by two factors: sampling time and the generation speed of the generative model. 
Although increasing the number of samples does raise the time required, it does not impose a significant burden and remains within an acceptable range. On average, our method takes less than 100 milliseconds to process each token.
On the WILD QA dataset, the Table \ref{tab:wild} shows that our method is on the same time scale as existing provably secure steganography algorithms and can be faster than the reordering distribution version of DISCOP.
On the social network XHS dataset in Table \ref{tab:XHS}, our method’s runtime is nearly identical to that of standard random sampling algorithms, and its time performance is comparable to existing provably secure algorithms such as METEOR and the non-reordering version of DISCOP, demonstrating the time efficiency of our approach.
Also, we observe that iMEC takes noticeably longer to generate stegotext. The original iMEC implementation was developed on GPT-2, whereas we reimplemented it for LLMs. Since the vocabularies of these models are much larger than that of GPT-2 (about 50K tokens), the time cost of running the MEC algorithm in our setting is substantially higher. This is consistent with the complexity of iMEC reported in Table~\ref{tab:summary}, namely $O(2^{B}\lvert V\rvert)$, where B is the block size and $\lvert V\rvert$ is the vocabulary size.

In the uniform distribution scenario, when the number of samples does not exceed $2^{14}$, our method achieves the same generation speed as random sampling. This trend is consistently observed across the LLAMA3, MISTRAL, and QWEN models.
Our implementation is in Python, so performance could be significantly optimized by replacing the sampling component with C/C++. This would bring the encoding speed closer to the model’s generation rate.

\subsubsection{On the Generation Quality of Stego}

In Tables 1 and 2, we present the experimental results on the XHS dataset and the QA dataset. (In the uniform distribution scenario, there is no need to discuss the quality of the generated text.) It is evident that, in the text generation task, the stegotext produced by our method exhibits linguistic quality similar to that of text generated through random sampling, both in terms of PPL and diversity metrics. This aligns with our security proof, demonstrating that the stegotext generated by our method is indistinguishable from the cover text.
In terms of fluency, our method achieved PPL scores similar to those of texts generated through random sampling. Higher PPL scores can sometimes result from increased entropy. In fact, studies have shown that blindly aiming for low PPL scores can lead to bland, redundant text or repetitive loops \cite{toppsampling}. However, the texts generated by our method did not exhibit these undesirable patterns and maintained good readability.

\subsubsection{Decoding Correctness}
\label{subsec:correctness}

We evaluate decoding correctness using the bit-level success rate (SR). As shown in Table~\ref{tab:wild} and Table~\ref{tab:XHS}, 
Our method achieves an SR of $1.0000$ in all settings, which indicates that the proposed adaptive dynamic sampling scheme yields deterministic, invertible decoding at the bit level in our experimental range. 
In contrast, iMEC exhibits noticeable instability in correctness. These results are consistent with the probabilistic nature of iMEC decoding.

\subsection{Examples of generated stegotexts}

We present demo stegotexts generated by existing provably secure generative steganography methods and our own steganography constructions by \textit{Llama3 8B} in Table \ref{casestudyWILD}. It can be seen that the stegotext generated by our scheme is nearly indistinguishable in quality from text produced by random sampling from the language model.
Also, as shown in Table~\ref{tab:example-chatgpt}, the ADS-generated stegotext produced with \textit{gpt-3.5-turbo-1106} is fluent and shows no noticeable anomalies. In this case, we can even regard the entire sentence as a single sampling outcome, where the paragraph corresponds to the concatenation of sampled steganographic sentences.

\begin{table*}[htbp]
    \centering
    \caption{Examples in WILD with \textsc{Llama3}}
     {
    \begin{tabularx}{\textwidth}{>{\raggedright\arraybackslash}p{3cm}|X} 
        \toprule[1.5pt]
        
        \textbf{Method} & \textbf{Generated Texts} \\
        \midrule
        Prompt & Please provide an action plan for reducing carbon footprint on a corporate level. \\
        \midrule
        Random Sampling & Here is a comprehensive action plan for reducing carbon footprint on a corporate level: 

        **I. Set Goals and Track Progress**

        1. Set specific, measurable, achievable, relevant, and time-bound (SMART) targets for reducing carbon footprint.
        
        2. Establish a carbon footprint tracking system to monitor progress and identify areas for improvement.
        
        3. Report carbon footprint reduction milestones to stakeholders, including employees, customers, and investors...\\
        \midrule
        METEOR \cite{meteor2021} & Here's a comprehensive action plan for reducing carbon footprint on a corporate level: 

        **I. Assessment and Goal-Setting**

        1. Conduct a carbon footprint assessment to identify areas of impact:
        
        - Energy consumption (electricity, gas, water)
        
        - Transportation (company-owned vehicles, employee travel)
        
        - Supply chain emissions (procurement, logistics)
        
        - Waste management (recycling, disposal)
        
        2. Set specific, measurable, achievable, relevant, and time-bound (SMART) targets... \\
        \midrule
        DISCOP \cite{ding2023discop} & Here is a comprehensive action plan for reducing carbon footprint on a corporate level: 

        **I. Conduct a Carbon Footprint Assessment**

        1. Engage a third-party auditor or consultant to conduct a comprehensive carbon footprint assessment of your organization's operations, including Scope 1 (direct emissions), Scope 2 (indirect emissions), and Scope 3 (value chain emissions).
        
        2. Identify key areas of energy consumption, such as:
        
        - Buildings and facilities
        
        - Transportation... \\
        \midrule

        
        
        SparSamp~\cite{wang2025sparsamp} &
       Here is an action plan for reducing carbon footprint on a corporate level:  **Short-term goals (0-12 months)**
       
       1. **Conduct a carbon footprint assessment**: Engage a third-party consultant to conduct a thorough assessment of the company's current carbon footprint, including energy consumption, transportation emissions, and waste management.
       
       2. **Set reduction targets**: Based on the assessment, set specific, measurable, achievable, relevant, and time-bound (SMART) targets for reducing the company's carbon footprint...
        \\\midrule
        iMEC~\cite{IMECde2022perfectly}&
        Action Plan for Reducing Carbon Footprint on a Corporate Level:
        
        **Step 1: Conduct an Initial Assessment (Weeks 1-4)**
        1. Identify the primary sources of carbon emissions within the organization:
        * Energy consumption (heating, cooling, lighting, etc.)
        * Transportation (company cars, fleet, employee commuting)
        * Travel (business flights, hotels, etc.)
        * Supply chain and operations (manufacturing, procurement, etc.)
        
        2. Gather data on current energy consumption, transportation, and travel practices...
        \\\midrule
        Ours & Reducing a corporation's carbon footprint is a crucial step towards sustainability and environmental responsibility. Here is a comprehensive action plan for corporate carbon footprint reduction: 

        **I. Conduct a Carbon Footprint Assessment**

        1. Identify the company's scope of activities: Determine what activities and products fall within the company's boundary for carbon footprint calculation.
        
        2. Gather data: Collect data on energy consumption, water usage, waste production, and greenhouse gas emissions from various sources...\\
        \bottomrule[1.5pt]
    \end{tabularx}}
    \label{casestudyWILD}
\end{table*}

\begin{table}[t]
  \centering
  \small
  \setlength{\tabcolsep}{6pt}
  \renewcommand{\arraystretch}{1.15}
  \begin{tabularx}{\linewidth}{@{} l L @{}}
    \toprule[1.5pt]

    \textbf{Prompt} & What is Steganography?\\

    \textbf{Stego text}  &Steganography is the practice of concealing a message, image, or file within another message, image, or file in such a way that the existence of the concealed information is hidden. It is often used to transmit secret information or to protect sensitive data from unauthorized access. Common techniques for steganography include hiding data within digital images, audio files, or even text documents, making it difficult for others to detect the hidden information without the proper tools or knowledge.\\\midrule
    \textbf{Prompt} & Give me a short introduction to large language model.\\
    \textbf{Stego text} &A large language model is a type of artificial intelligence that has been trained on vast amounts of text data in order to understand and generate human-like language. These models use advanced machine learning algorithms to analyze and predict sequences of words, enabling them to comprehend...\\
\midrule
    \textbf{Prompt} & 1+1=?\\
    \textbf{Stego text} & 1+1 equals 2.\\

    \bottomrule[1.5pt]
  \end{tabularx}
  \caption{Examples of stegotext generated by \texttt{gpt-3.5-turbo-1106}.}
  \label{tab:example-chatgpt}
\end{table}

\section{Limitations}
\textbf{Limitation of APIs.} Our steganography method require a model API that accepts a seed to control its randomness. This requirement can be met in practice: several major LLM e.g., Qwen-max, Qwen-Plus, gpt-4-1106-preview, gpt-3.5-turbo-1106) already expose a seed parameter, and popular image generation systems such as Midjourney and Stable Diffusion also supports such APIs.
However, the availability of seed control are ultimately governed by each API provider’s evolving product and security policies. For example, to the best of our knowledge, OpenAI, Qwen, and other existing LLM providers currently state that it will make a best effort to sample deterministically for fixed seeds and parameters, but also explicitly note that determinism is not strictly guaranteed. This implies that API-level seed control may be unstable over time or across deployments, which in turn may limit the long-term robustness of our scheme when used purely with third-party black-box services. Users of our steganography scheme should pay close attention to any changes in the model API provider’s policies and support for seed control.
In our view, the most reliable deployment scenario for the proposed steganography method is one in which the user hosts or controls their own model API (e.g., on a private or cloud-managed server), so that seed determinism and sampling behavior can be ensured. In such settings, our approach is also fully backward-compatible with open-source white-box language models, where both the random seed and the sampling routine can be specified precisely.

\textbf{Limitation of applying our method to image/video models.} Although it is feasible to apply our method to image or video models as long as they support seed-based sampling, we note that generating images or videos is significantly more computationally expensive than generating text. It may take several minute to generate an image sample, which is not practical in real-world scenario. 
Our method is best suited for existing autoregressive language models, and we hope to further investigate its application in real-world image and video settings with higher efficiency.

\textbf{Limitation of sampling speed.}
Since this work does not specifically target sampling acceleration, our current implementation, as analyzed in the Time Efficiency subsection, requires approximately 0.02–0.04 seconds to generate each stego token when the number of samples per step is set to $2^{14}$. As the sampling budget increases, the generation time per stego token naturally grows as well, so excessively large choices of $N$ can be expected to introduce noticeable computational overhead. We view the design of more efficient sampling strategies or dedicated steganographic algorithms aimed at improving time efficiency as an important direction for future research. We are also considering converting the code to a C/C++ implementation.

\section{Conclusion}

In this paper, we propose a provably secure steganography scheme built on Adaptive Dynamic Sampling, which samples from model API with seed without requiring knowledge of the explicit model distribution. We design a strategy to determine the number of samples at each step and embed secret messages into the sampling indices without altering the model’s original generation process, ensuring theoretical indistinguishability between stego and cover. Extensive evaluations on three popular LLMs and three benchmark datasets show that our scheme achieves high generative quality, efficiency, and capacity, comparable to provably secure methods that require an explicit distribution.

\section{Acknowledgments}
This work was supported by the Natural Science Foundation
of China under Grant U2336208.

\section{Ethical Considerations}
We propose a provably secure steganographic scheme that embeds secret messages into seemingly innocuous carriers, requiring no access to the model’s explicit distribution and relying on controlled sampling access to the model.
Consistent with prior work~\cite{meteor2021,ding2023discop,wang2025sparsamp}, we treat the secret information as a random binary bitstream in our research. Given the existence of provably secure steganography constructions, we do not expect our contribution to materially exacerbate existing ethical concerns. Nonetheless, we provide a stakeholder and dual-use risk analysis, as well as a discussion of potential mitigation strategies, in the hope that these reflections may be of some use for future work on steganography.
\subsection{Stakeholder and Dual-Use Risk Analysis}
Steganography application context centers on two primary stakeholder groups: steganography users and monitors. For steganography users (both sender and receiver), the goal is to transmit information covertly. These users fall into two broad scenarios. First are legitimate users, for example, journalists or ordinary citizens in environments of excessive surveillance, who may rely on steganography to safeguard privacy and exercise free expression. The second comprises malicious users who seek harmful coordination or the concealment of illegal content. In our study, we abstract away the secret’s semantics and treat the payload as a random binary bitstream. In real deployments, we acknowledge the dual-use risk that adversaries could embed harmful or illegal secret messages. The same properties that protect vulnerable users can, if misappropriated, also facilitate abuse.

For monitors, we distinguish between illegal monitors and legitimate defenders (trust-and-safety teams and forensic analysts). Illegal monitors conduct unauthorized inspection of platforms and communications, raising serious privacy and due-process concerns. Legitimate defenders protect platform security by reviewing user content and leveraging behavior-aware signals (such as account history and interaction patterns) to assess risk and infer user intent, including for highly covert communications. Our steganography work is motivated by the goal of resisting illegal monitors.

\subsection{Mitigating Potential Harm}
Our contribution is a theoretical prototype for scientific study, not a production-ready system. To limit potential harm, all experiments are conducted offline in sandboxed environments; no stegotext is posted to live networks. Any released code is research-use only: will not provide operational keys or publish artifacts that would materially elevate adversarial use beyond what is already possible with existing steganography schemes~\cite{meteor2021,ding2023discop,wang2025sparsamp}. We will also align the work with defensive practice. Alongside the steganography code, we will provide evaluation protocols and steganalysis baselines in real-world scenarios, and we explicitly clarify that provable indistinguishability applies to the cover distribution, not to user or network behavior. Consequently, behavior-based defenses remain applicable in practice. Finally, we include prominent usage warnings: the code targets theoretical research, and \textbf{steganography is not a panacea that takes the risk of circumvention away.} It can reduce content-level detectability but cannot eliminate operational or network risks. We strongly recommend that users comply with all applicable local laws and regulations.

\section{Open Science}
Our steganography algorithms and prompts are available at Zenodo, \url{https://zenodo.org/records/18152848}. 
The project includes: the ADS steganography encoding and decoding algorithms, the prompts used in our experiments, and the Python environment dependencies.


\newpage
\newpage
\bibliographystyle{unsrt}
\bibliography{neurips_2024P}

\end{document}